\documentclass[twocolumn,showpacs,preprintnumbers,amsmath,amssymb]{revtex4}
\usepackage{graphicx}
\usepackage{dcolumn}
\usepackage{bm}

\begin{document}

\title{Probing optomechanical correlations between two optical beams\\
down to the quantum level}

\author{P. Verlot}
\author{A. Tavernarakis}
\author{T. Briant}
\author{P.-F. Cohadon}
\author{A. Heidmann}
\affiliation{Laboratoire Kastler Brossel, UPMC-ENS-CNRS, Case 74, 4
place Jussieu, F75252 Paris Cedex 05, France}

\date{\today}

\begin{abstract}
Quantum effects of radiation pressure are expected to limit the
sensitivity of second-generation gravitational-wave
interferometers. Though ubiquitous, such effects are so weak that
they haven't been experimentally demonstrated yet. Using a
high-finesse optical cavity and a classical intensity noise, we
have demonstrated radiation-pressure induced correlations between
two optical beams sent into the same moving mirror cavity. Our
scheme can be extended down to the quantum level and has
applications both in high-sensitivity measurements and in quantum
optics.
\end{abstract}


\maketitle

Quantum effects of optomechanical coupling, the radiation-pressure
coupling between a moving mirror and an incident light field, were
first studied in the framework of gravitational-wave detection
\cite{Braginsky92,Bradaschia90,Abramovici92}, enforcing quantum
limits to the sensitivity of large-scale interferometers
\cite{Caves81,Jaekel90,AdvLIGO}. Overcoming these limits was a
major motivation for the quantum optics experiments performed
shortly after, such as squeezing of the light field
\cite{Walls,BellLabs} or quantum non demolition (QND) measurements
\cite{GrangierNature,QND,GrangierQND}. Such pioneering experiments
were performed with nonlinear optical media, but optomechanical
coupling was soon proposed as a candidate nonlinear mechanism of
its own \cite{Fabre94,Jacobs94,Heidmann97}, based upon
correlations between light intensity and mirror displacement
induced by radiation pressure.

\begin{figure}
\includegraphics[width=7 cm]{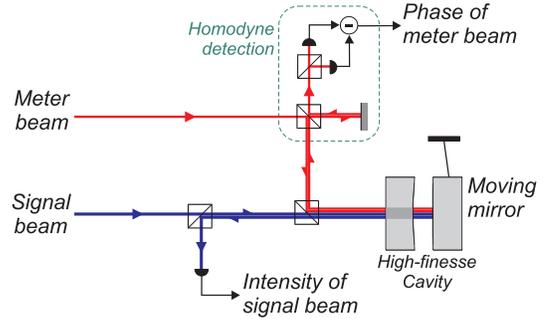}
\caption{Principle of the direct observation of optomechanical
correlations. Both an intense signal beam and a weaker meter beam
are sent into a resonant high-finesse cavity with a moving mirror.
Intensity fluctuations of the signal beam are imprinted by
radiation pressure onto the position fluctuations of the moving
mirror, and subsequently onto the phase fluctuations of the meter
beam. The two reflected beams then display intensity-phase
correlations, retrieved with both a photodiode and a homodyne
detection.} \label{fig:QND}
\end{figure}

The first experiments fell short of the quantum regime
\cite{EurophysLett,LaserCool,Tittonen} and even though recent ones
demonstrated a much larger optomechanical coupling \cite{PRL-LKB,
NatureLKB, Nature-Sylvain, PRL_Mavalvala, Kippenberg,
Nature-Harris}, they mainly focussed on the possible demonstration
of the quantum ground state of a mechanical resonator
\cite{Cleland,Schwab}. To observe the optomechanical correlations,
two beams have to be sent upon the moving mirror (see Fig.
\ref{fig:QND}): the intensity fluctuations of the first, intense,
signal beam drive the mirror into motion by radiation pressure,
whereas the resulting position fluctuations are monitored through
the phase of the second, weaker, meter beam. As the intensity
fluctuations of the signal beam are unaltered by reflection upon
the mirror and as far as the radiation pressure of the meter beam
is negligible, the intensity-phase correlations observable between
the two reflected beams provide a direct measurement of the
optomechanical correlations.

To monitor these radiation-pressure effects down to the quantum
level and hence perform a real-time QND measurement of the signal
intensity via the meter phase \cite{Heidmann97}, one has first to
enhance the optomechanical coupling by using a high-finesse cavity
with a moving mirror, as shown in Fig. \ref{fig:QND}. The position
fluctuations $\delta x_{\rm rad}$ induced by the quantum intensity
fluctuations of the signal beam also have to be the dominant noise
source, which requires to lower the thermal fluctuations $\delta
x_{\rm T}$ of the moving mirror. For a harmonic oscillator of mass
$M$, resonance frequency $\Omega_{\rm M}/2\pi$, and mechanical
quality factor $Q$, the corresponding ratio between the radiation
pressure and thermal noise spectra can be written
\cite{Heidmann97}
\begin{eqnarray}
\frac{S_x^{\rm rad}}{S_x^{\rm T}}&\simeq& 2.3 \left(\frac{\cal F}{
300\,000}\right)^2 \left(\frac{800\,{\rm nm}}{\lambda}\right)
\left(\frac{P_{\rm in}}{1\,{\rm mW}}\right) \nonumber \\
&&\times \left(\frac{1\,{\rm mg}}{M}\right)
\left(\frac{Q}{10^6}\right) \left(\frac{1\,{\rm MHz}}{\Omega_{\rm
M}/2\pi}\right) \left(\frac{1\,{\rm
K}}{T}\right)\label{eq:ratio_rad_T}
\end{eqnarray}
where $T$ is the environment temperature, $\mathcal{F}$ the cavity
finesse, $\lambda$ the optical wavelength, and $P_{\rm in}$ the
incident intensity of the signal beam. The stated values have all
already been achieved independently in various state-of-the-art
optomechanical systems \cite{PRL-LKB, NatureLKB, Nature-Sylvain,
PRL_Mavalvala, Kippenberg, Nature-Harris, Rugar, Caniard}, but
combining the favourable mechanical behaviour of NEMS \cite{Rugar}
with a very high optical finesse \cite{Caniard} is an even greater
experimental challenge.

In this work, we report the observation of optomechanical
correlations measured close to the quantum level. To reach a ratio
(\ref{eq:ratio_rad_T}) as large as possible, we favour the optical
characteristics and use a fused silica moving mirror, which
provides both a very high optical finesse \cite{Caniard} and
mechanical quality factor \cite{PRA_modes_gaussiens}, at the
expense of a larger mass. The optomechanical correlations have
then been measured with a tiny classical intensity modulation of
the signal beam that mimics at a higher level its quantum
fluctuations \cite{Caniard,PRL_Australiens_LQS}.

\begin{figure}
\includegraphics[width=8.6 cm]{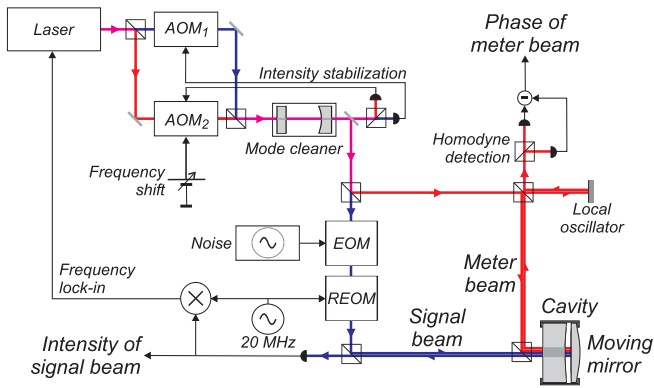}
\caption{Experimental setup. The laser beam is split in two
orthogonally polarized beams, which are both sent into the moving
mirror cavity. A resonant electro-optical modulator (REOM) is used
to lock the laser onto the optical resonance via a
Pound-Drever-Hall technique. The residual birefringence of the
cavity is compensated by the frequency shift of two acousto-optic
modulators (AOM), also used to stabilize the intensities of both
beams after their spatial filtering by the mode cleaner cavity. A
second EOM modulates the intensity of the signal beam to mimic
quantum radiation-pressure noise. Intensity fluctuations of the
reflected signal beam are monitored with a photodiode, as are the
phase fluctuations of the meter beam with a quantum-limited
homodyne detection. For simplicity, most polarizing elements are
not shown.} \label{fig:Setup}
\end{figure}

Our experimental setup is based on a single-ended optical cavity,
with a 1-inch fused silica cylindrical input mirror. The moving
mirror, used as end mirror, is a plano-convex 34-mm diameter and
2.5-mm thick mirror, which displays gaussian internal vibration
modes \cite{PRA_modes_gaussiens}. We work at frequencies close to
a mechanical resonance with the following optomechanical
characteristics, deduced from the thermal noise spectrum at room
temperature: $\Omega_{\rm M}/2\pi= 1.125\,{\rm MHz}$, $M=500\,{\rm
mg}$, $Q=500\,000$.

The low roughness of the silica substrates allows for optical
coatings with very low losses: we have obtained a cavity finesse
$\mathcal{F}=330\,000$, mainly limited by the 20-ppm transmission
of the input mirror. This is crucial for quantum optics
experiments for which loss has to be avoided to get large
correlations between intracavity and reflected fields. We use a
short, 0.33-mm long, cavity in order to keep a sufficient cavity
bandwidth ($\Omega_{\rm cav}/2\pi=700\,{\rm kHz}$) and to prevent
laser frequency noise from limiting the displacement sensitivity.
The cavity is operated in vacuum to increase the mechanical
quality factors.

The cross-polarized signal and meter beams entering the cavity are
provided by a Ti:Sa laser working at 810 nm. As the cavity is
birefringent (with a 5-MHz frequency mismatch between the two
optical resonances), two acousto-optic modulators (AOM in Fig.
\ref{fig:Setup}) independently detune the two beams so that they
both match the cavity resonance. The overall resonance is
controlled by locking the laser frequency via a Pound-Drever-Hall
technique: the incident signal beam is phase-modulated at 20\,MHz
by a resonant electro-optical modulator (REOM), and the resulting
intensity modulation of the reflected beam provides the error
signal. A mode cleaner cavity filters potential degradations of
the spatial profile of both beams, while their intensities after
the mode cleaner are stabilized by a servo-loop which drives the
amplitude control of the AOMs.

The phase fluctuations $\delta\varphi_m^{\rm out}(t)$ of the
reflected meter beam are monitored by a homodyne detection, with a
local oscillator derived from the incident meter beam and
phase-locked in order to detect the phase quadrature. For an
incident power of 50\,$\mu$W, one gets a shot-noise-limited
displacement sensitivity of $2.7\times 10^{-20}\,{\rm
m}/{\sqrt{\rm Hz}}$ at frequencies above 200\,kHz. Intensity
fluctuations $\delta I_s^{\rm out}(t)$ of the reflected signal
beam are monitored by a high-efficiency photodiode. We have
carefully eliminated unwanted optical reflections so that the
optical rejection of the double-beam system is higher than 35 dB:
the phase fluctuations of the meter beam are insulated from the
intensity fluctuations of the signal beam in such a way that
observable effects of the signal beam are necessarily induced by
intracavity radiation pressure.

In order to mimic the quantum fluctuations of radiation pressure,
the signal beam is intensity-modulated with an electro-optic
modulator (EOM) before entering the cavity to produce a classical
intracavity radiation-pressure noise
\cite{Caniard,PRL_Australiens_LQS}. The digitized driving noise is
centered at a frequency $\Omega_{\rm c}$ close to the mechanical
resonance frequency $\Omega_{\rm M}$, and has a typical bandwidth
of a few hundreds of Hz, larger than any bandwidth used in the
correlations acquisition process. To generate a gaussian intensity
noise of the form $\delta I_s^{\rm in}(t)=A(t)
\cos\left(\Omega_{\rm c}t+\varphi(t)\right)$ where $A(t)$ is a
random function with a gaussian distribution around 0 and
$\varphi(t)$ a randomly-distributed phase, we decompose the noise
into its quadratures \cite{EPJD}:
\begin{equation}
\delta I_s^{\rm in}(t)=X_{I_s}^{\rm in}(t)\cos\left(\Omega_{\rm c}
t\right)+Y_{I_s}^{\rm in}(t)\sin\left(\Omega_{\rm c}
t\right).\label{eq:def_Quad}
\end{equation}
The quadratures are produced from a dual-channel arbitrary
waveform generator Tektronix AFG3022B, and then summed to drive
the EOM. The slowly-varying gaussian noise functions $X_{I_s}^{\rm
in}(t)$ and $Y_{I_s}^{\rm in}(t)$ are randomly generated by a
computer and loaded into the generator as amplitude arrays.

The experiment is performed as follows. Both optical beams are
locked onto the resonance of the cavity, with incident powers
$P_s^{\rm in}= 150\,\mu{\rm W}$ for the signal beam and $P_m^{\rm
in}= 500\,\mu{\rm W}$ for the meter. The EOM drives a classical
radiation-pressure noise with an amplitude level as compared to
thermal noise of $\sqrt{S_x^{\rm rad}/S_x^{\rm T}}\simeq 5$, and
with a center frequency $\Omega_{\rm c}/2\pi = 1.123\,{\rm MHz}$,
about 600 mechanical linewidths below the mechanical resonance.
The experimental signals are independently acquired by two
spectrum analyzers Agilent MXA set in I/Q mode in order to
directly extract the quadratures $X_{I_s}^{\rm out}(t)$,
$Y_{I_s}^{\rm out}(t)$, $X_{\varphi_m}^{\rm out}(t)$, and
$Y_{\varphi_m}^{\rm out}(t)$ of the reflected signal intensity and
meter phase, respectively. Both analyzers are locked at the same
central frequency $\Omega_{\rm c}$ with an analysis bandwith of
$400\,{\rm Hz}$, and synchronously triggered with the waveform
generator. Temporal evolution of the quadratures are then acquired
over a span time of $200\,{\rm ms}$, equal to the scan time of the
digitized amplitude-modulation arrays of the generator.

\begin{figure}
\includegraphics[width=8.5 cm]{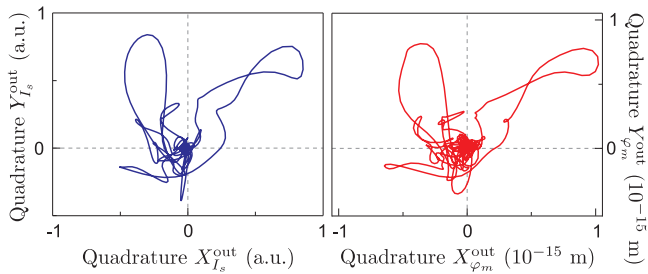}
\caption{Phase-space trajectories of the intensity noise of the
signal beam (left) and the phase noise of the meter beam (right),
in the case $\sqrt{S_x^{\rm rad}/S_x^{\rm T}}\simeq 5$. The phase
noise is calibrated as displacements of the moving mirror.}
\label{fig:Phases}
\end{figure}

Fig. \ref{fig:Phases} presents the observed phase-space
trajectories: clear correlations are evident between the intensity
noise of the signal beam (left) and the meter phase noise (right).
Neglecting optical losses and irrelevant noises such as the
quantum fluctuations of the meter beam, this can be interpreted
from the following input-output relations for the fluctuations at
frequency $\Omega_{\rm c}$ \cite{Heidmann97}:
\begin{eqnarray}
\delta I_s^{\rm out}[\Omega_{\rm c}]&=&\frac{1+i\omega_{\rm
c}}{1-i\omega_{\rm c}}\delta I_s^{\rm in}[\Omega_{\rm
c}],\label{eq:Iout}\\
\delta \varphi_m^{\rm out}[\Omega_{\rm c}]&=&\frac{8{\cal
F}}{\lambda(1-i\omega_{\rm c})}\delta x[\Omega_{\rm
c}],\label{eq:Phiout}
\end{eqnarray}
where $\omega_{\rm c}=\Omega_{\rm c}/\Omega_{\rm cav}$, and
$\delta x=\delta x_{\rm T}+\delta x_{\rm rad}$ is the mirror
motion, including the thermal noise and the radiation-pressure
noise given by
\begin{equation}
\delta x_{\rm rad}[\Omega_{\rm c}]=\frac{8 {\cal F}}{\lambda
(1-i\omega_{\rm c})}\hbar \chi[\Omega_{\rm c}] \delta I_s^{\rm
in}[\Omega_{\rm c}],\label{eq:dxrad}
\end{equation}
where $\chi[\Omega_{\rm c}]$ is the mechanical susceptibility of
the moving mirror. The reflected signal intensity noise reproduces
the incident one, with a global phase shift depending on
$\omega_{\rm c}$ [eq. (\ref{eq:Iout})], whereas the reflected
meter phase reproduces the incident signal intensity $\delta
I_s^{\rm in}$ via the mirror motion [eqs. (\ref{eq:Phiout}) and
(\ref{eq:dxrad})]. It is superimposed to the thermal noise $\delta
x_{\rm T}$ of the mirror which is responsible for the small
differences observed between the two phase-space evolutions in
Fig. \ref{fig:Phases}. Other noises such as the quantum phase
noise of the incident meter beam, which limits the sensitivity of
the displacement measurement, are negligible in our current setup
with a level at least $15\,{\rm dB}$ below the thermal noise. Also
note that the meter phase in Fig. \ref{fig:Phases} is calibrated
in terms of the equivalent displacements of the moving mirror,
with a typical level at $10^{-15}\,{\rm m}$, and the curve has
been rotated in phase space in order to compensate for the global
phase shifts due to $\omega_c$ and to the mechanical response
$\chi[\Omega_{\rm c}]$.

We have obtained similar results with a center frequency
$\Omega_{\rm c}$ closer or equal to the mechanical resonance
frequency. In that case, the resonance amplifies the
radiation-pressure and thermal displacements by a factor up to the
quality factor $Q$, but the phase shift of the mechanical response
across the resonance frequency has to be taken into account to
deconvolve the observed data. We focus in the following on
experimental results obtained at low frequency.

The results can be made more quantitative by computing the
correlation coefficient $C_{I_s,\varphi_m}$ defined from the two
trajectories in phase-space:
\begin{equation}
C_{I_s,\varphi_m}=\frac{\left|\langle\delta I_s^{\rm out}\,
{\delta \varphi_m^{\rm out}}^\star\rangle\right|^2}
{\langle\left|\delta I_s^{\rm out}\right|^2\rangle\,
\langle\left|\delta \varphi_m^{\rm
out}\right|^2\rangle},\label{eq:correl}
\end{equation}
where the brackets $\langle ... \rangle$ stand for a temporal
average. We obtain a coefficient $C_{I_s,\varphi_m}\simeq 0.96$
for the data presented on Fig. \ref{fig:Phases}, in perfect
agreement with the value $\left(1+S_x^{\rm T}/S_x^{\rm
rad}\right)^{-1}$ deduced from Eqs. (\ref{eq:Iout}) to
(\ref{eq:dxrad}).

\begin{figure}
\includegraphics[width=8.5 cm]{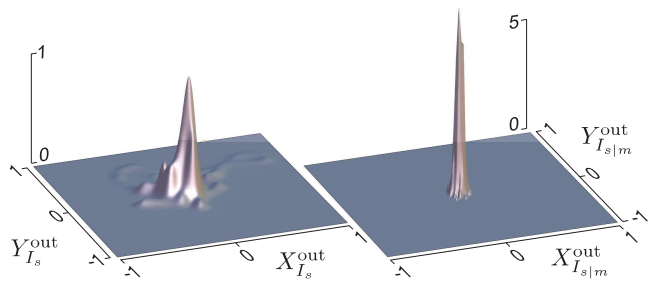}
\caption{Probability distributions in phase space of the signal
intensity fluctuations $\delta I_s^{\rm out}$ (left) and of the
conditional fluctuations $\delta I_{s|m}$ deduced from the meter
measurement (right). Note the sharper peak, related to the lower
conditional variance, and the factor 5 between the two vertical
scales.} \label{fig:Histogrammes}
\end{figure}

As in usual QND measurements \cite{GrangierNature}, optomechanical
correlations can also be quantified by the knowledge we have on
the signal intensity from the measurement of the meter phase. The
resulting distribution is given by the conditional fluctuations
\begin{equation}
\delta I_{s|m}=\delta I_s^{\rm out}-\frac{\langle\delta I_s^{\rm
out}\, {\delta \varphi_m^{\rm
out}}^\star\rangle}{\langle\left|\delta \varphi_m^{\rm
out}\right|^2\rangle}\delta \varphi_m^{\rm out}.\label{eq:Icond}
\end{equation}
Figure \ref{fig:Histogrammes} presents the respective probability
distributions in phase space for the uncorrected intensity
fluctuations $\delta I_s^{\rm out}$ and the conditional ones
$\delta I_{s|m}$, obtained as normalized histograms of the data of
Fig. \ref{fig:Phases}. The shrinking of the distribution is
related to the lower conditional dispersion, reduced by a factor
$\simeq 5$, as can be deduced from Eqs. (\ref{eq:Iout}) to
(\ref{eq:Icond}):
\begin{equation}
\Delta I_{s|m} = \sqrt{1-C_{I_s,\varphi_m}} \,\Delta I_s^{\rm
out}\simeq 0.2\,\Delta I_s^{\rm out}.\label{eq:VarianceCond}
\end{equation}

Our experimental setup enables to demonstrate optomechanical
correlations even in the case of radiation-pressure effects
smaller than the thermal noise $\left(S_x^{\rm rad}\ll S_x^{\rm
T}\right)$. In such a case, as the reflected meter phase
fluctuations $\delta\varphi_m^{\rm out}$ are mainly related to
random thermal noise, the correlation coefficient deduced from the
temporal average of a single 200-ms run has little meaning, and
experimental values fluctuate from one run to the other.
Nevertheless, repeating such runs and averaging all these
experimental outcomes eventually yields a steady value.

\begin{figure}
\includegraphics[width=6.5 cm]{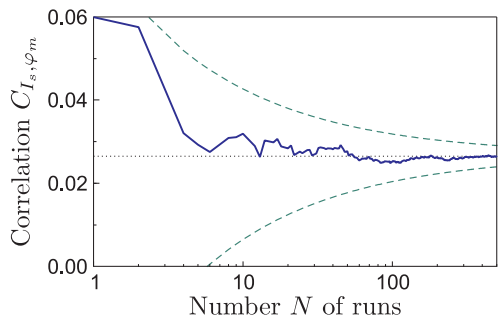}
\caption{Estimated optomechanical correlation coefficient
$C_{I_s,\varphi_m}$ with respect to the number $N$ of independent
200-ms runs averaged. Dashed lines delineate the expected
$1/\sqrt{N}$ statistical uncertainty region.}
\label{fig:Moyennage}
\end{figure}

Fig. \ref{fig:Moyennage} presents the estimate of the correlation
coefficient obtained with $S_x^{\rm rad}/S_x^{\rm T}\simeq 0.03$,
as a function of the number $N$ of runs averaged, up to $N=500$.
The resulting correlation coefficient tends to its small but
non-zero expected value $\left(1+S_x^{\rm T}/S_x^{\rm
rad}\right)^{-1}\simeq 0.03$, with a statistical uncertainty at
least 10-times smaller ($2.5\times 10^{-3}$ for 500 averages).

We have thus demonstrated optomechanical correlations between two
light beams. Such correlations are still at the classical level
but we note that for our system, $S_x^{\rm rad}/S_x^{\rm T}\simeq
10^{-3}$ for quantum noise and a temperature of $1\,{\rm K}$.
Averaging the experimental signal once working at low temperature
should enable to retrieve the corresponding quantum correlations
and hence demonstrate radiation-pressure noise, which is expected
to be a severe limitation of second-generation gravitational-wave
interferometers \cite{AdvLIGO}. With an upgrade of our
experimental setup, one can also envision radiation-pressure
induced quantum optics experiments, such as optomechanical
squeezing \cite{Fabre94} or QND measurements \cite{Jacobs94,
Heidmann97}.

We gratefully acknowledge the Laboratoire des Mat\'eriaux
Avanc\'es for the low-loss optical coating of the mirrors. We are
grateful to Matteo Barsuglia and Eric Chassande-Mottin for
fruitful discussions. This work was partially funded by EGO
(collaboration convention EGO-DIR-150/2003 for a study of quantum
noises in gravitational wave interferometers) and by the
Integrated Large Infrastructures for Astroparticle Science (ILIAS)
of the Sixth Framework Program of the European Community.

%%%%%%%%%%%%%%%%%%%%%%%%%%%%%%%%%%%%%%%%%%%%%%%%%%%%%%%%%%%%%


\begin{thebibliography} {0}

\bibitem{Braginsky92}   V.B. Braginsky and F. Ya Khalili, {\it Quantum
Measurement} (Cambridge University Press, 1992).

\bibitem{Bradaschia90} C.~Bradaschia {\it et al.}, Nucl.
Instrum. Meth. A {\bf 289}, 518 (1990).

\bibitem{Abramovici92} A. Abramovici {\it et al.}, Science
{\bf 256}, 325 (1992).

\bibitem{Caves81} C.M. Caves, Phys. Rev. D {\bf 23}, 1693
(1981).

\bibitem{Jaekel90} M.T. Jaekel and S. Reynaud, Europhys.
Lett. {\bf 13}, 301 (1990).

\bibitem{AdvLIGO} P. Fritschel, Proc. SPIE {\bf 4856}, 282
(2003).

\bibitem{Walls} D.F. Walls, Nature (London) {\bf 306}, 141 (1983).

\bibitem{BellLabs} R.E. Slusher, L.W. Hollberg, B. Yurke, J.C.
Mertz, and J.F. Valley, Phys. Rev. Lett. {\bf 55}, 2409 (1985).

\bibitem{GrangierNature} P. Grangier, J.A. Levenson, J.-P.
Poizat, Nature (London) {\bf 396}, 537 (1998).

\bibitem{QND} N. Imoto, H.A. Haus, and Y. Yamamoto, Phys. Rev. A
{\bf 32}, 2287 (1985).

\bibitem{GrangierQND} J.-F. Roch {\it et al.}, Phys. Rev. Lett.
{\bf 78}, 634 (1997).

\bibitem{Fabre94} C. Fabre {\it et al.}, Phys. Rev. A {\bf
49}, 1337 (1994).

\bibitem{Jacobs94} K. Jacobs, P. Tombesi, M.J. Collett, and D.F. Walls,
Phys. Rev. A {\bf 49}, 1961 (1994).

\bibitem{Heidmann97} A. Heidmann, Y. Hadjar, and M.
Pinard, Appl. Phys. B {\bf 64}, 173 (1997).

\bibitem{EurophysLett} Y.~Hadjar, P.-F. Cohadon, C.G. Aminoff, M. Pinard, and A. Heidmann,  Europhys.
Lett. {\bf 47}, 545 (1999).

\bibitem{LaserCool} P.-F. Cohadon, A. Heidmann, and M. Pinard,
Phys. Rev. Lett. {\bf 83}, 3174 (1999).

\bibitem{Tittonen} I.~Tittonen {\it et al.}, Phys. Rev. A {\bf 59}, 1038 (1999).

\bibitem{PRL-LKB} O. Arcizet {\it et al.}, Phys. Rev. Lett. {\bf 97}, 133601
(2006).

\bibitem{NatureLKB} O. Arcizet, P.-F. Cohadon, T. Briant, M.
Pinard, and A. Heidmann, Nature (London) {\bf 444}, 71 (2006).

\bibitem{Nature-Sylvain} S. Gigan {\it et al.}, Nature (London) {\bf
444}, 67 (2006).


\bibitem{PRL_Mavalvala} T. Corbitt {\it et al.}, Phys. Rev.
Lett. {\bf 99}, 160801 (2007).

\bibitem{Nature-Harris} J.D. Thompson {\it et al.}, Nature (London) {\bf 452}, 72
(2008).

\bibitem{Kippenberg} A. Schliesser, O. Rivi\`ere, G. Anetsberger, O. Arcizet,
and T.J. Kippenberg, Nature Physics {\bf 4}, 415 (2008).

\bibitem{Cleland} R.G.~Knobel and A.N.~Cleland, Nature (London) {\bf
424}, 291 (2003).

\bibitem{Schwab} M.D. LaHaye, O. Buu, B. Camarota, and K.C. Schwab, Science {\bf
304}, 74 (2004).

\bibitem{Rugar} D. Rugar, R. Budakian, H.J. Mamin, and B.W. Chui,
Nature (London) {\bf 430}, 329 (2004).

\bibitem{Caniard} T. Caniard, P. Verlot, T. Briant, P.-F. Cohadon, and A.~Heidmann, Phys. Rev. Lett. {\bf 99}, 110801
(2007).

\bibitem{PRA_modes_gaussiens} T. Briant, P.-F. Cohadon, A.
Heidmann, and M. Pinard, Phys. Rev. A {\bf 68}, 033823 (2003).

\bibitem{PRL_Australiens_LQS}  C.M. Mow-Lowry, B.S. Sheard, M.B. Gray,
D.E.~McClelland, and S.E. Whitcomb, Phys. Rev. Lett. {\bf 92},
161102 (2004).

\bibitem{EPJD} T. Briant, P.-F. Cohadon, M. Pinard, and A. Heidmann,
Eur. Phys. J. D {\bf 22}, 131 (2003).

\end{thebibliography}
\end{document}